\documentstyle[12pt]{article}
\parskip=\medskipamount

\newcommand{\be}{\begin{equation}}
\newcommand{\ee}{\end{equation}}
\newcommand{\bea}{\begin{eqnarray}}
\newcommand{\eea}{\end{eqnarray}}
\newcommand{\nn}{\nonumber}
\newcommand{\p}[1]{(\ref{#1})}
\begin{document}
\begin{titlepage}

\begin{flushright}
JINR E2-98-55 \\
\tt{hep-th/9803176} \\
\end{flushright}

\begin{center}
{\large\bf Central Charge as the Origin of Holomorphic
Effective Action \\
in N=2 Gauge Theory}
\end{center}
\vspace{5mm}

\begin{center}
E. I. Buchbinder$\,^{a,b)}$, I. L. Buchbinder$\,^{c)}$,
E. A. Ivanov$\,^{b)}$
and S. M. Kuzenko\, ${}^{d)}$
\vspace{3mm}

${}^{a)}$ \footnotesize{{\it Department of Quantum Field Theory,
Tomsk State University\\
Tomsk 634050, Russia}}\\
\vspace{3mm}

${}^{b)}$ \footnotesize{{\it Bogoliubov Laboratory of Theoretical
Physics, Joint Institute of Nuclear Research \\
Dubna 141980, Russia}}\\
\vspace{3mm}

${}^{c)}$ \footnotesize{{\it Department of Theoretical Physics,
Tomsk State Pedagogical University\\
Tomsk 634041, Russia}}\\
\vspace{3mm}

${}^{d)}$ \footnotesize{{\it Department of Theoretical Physics,
Tomsk State University\\
Tomsk 634050, Russia}}  \\

\end{center}

\begin{abstract}
We explicitly demonstrate that the perturbative holomorphic
contribution to the off-shell effective action of $N=2$
$U(1)$ gauge supermultiplet is an entire effect of the minimal
coupling to a hypermultiplet with the mass generated by
a central charge in $N=2$ superalgebra. The central charge
is induced by a constant vacuum $N=2$ gauge superfield
strength spontaneously breaking the automorphism $U(1)_R$ symmetry of
$N=2$ superalgebra. We use the manifestly off-shell supersymmetric
harmonic superspace techniques of quantum calculations with the
central charge-massive hypermultiplet propagator.
\end{abstract}

\end{titlepage}

\newpage
\setcounter{page}{1}

\noindent
The holomorphy is one of the basic features of the
low-energy effective actions in supersymmetric field
theories (see, e.g., \cite{1} for a review). It means that
in supersymmetric models with complex superfields defined
on suitable complex subspace of the full superspace
some quantum corrections to effective action
frequently arise in the form of holomorphic
functions of these superfields integrated over the relevant
complex subspace. A good example of holomorphy in $N=1$
supersymmetric field theory is chiral effective potential \cite{2}.
Another example is the low-energy effective action of $N=2$
gauge theory where the term leading in momenta can be written as a
chiral $N=2$ superspace integral of a holomorphic function of the
chiral $N=2$ gauge superfield strength \cite{hol,seiberg}.  Just the
holomorphy of effective action together with an idea of duality allowed
Seiberg and Witten \cite{3} to find the exact low-energy effective action
in $N=2$ $SU(2)$ super Yang-Mills theory, with all non-perturbative
corrections taken into account. This result
entailed a plenty of works where various aspects of
the effective action in $N=2$ super Yang-Mills theories,
with and without matter, were examined.

In refs. \cite{4}, a perturbative approach to the effective action
in $N=2$ SUSY models was formulated in the $N=1$ superfield notation.
Although such a formalism allows us to obtain
the correct results, it does not possess a manifest $N=2$  supersymmetry.
This could (and does) lead to a number of technical problems,
as well as troubles with interpretation of the results.

Recently, in refs. \cite{5,7,6} a new manifestly $N=2$ supersymmetric
approach to the effective action
in $N=2$ supersymmetric field theories was
developed. It is based on the formulation of $N=2$
models in harmonic superspace \cite{8}. It was demonstrated that
the harmonic superspace provides an adequate arena for investigating
the classical and quantum aspects of theories with $N=2$
supersymmetry. This approach allows one both to analyze
general properties of the
effective action and to perform actual calculations.

The problem of deriving the holomorphic corrections to the effective
action of $N=2$ $U(1)$ gauge superfield minimally coupled to
the charged matter hypermultiplet was treated by the
harmonic superspace methods in our paper \cite{5}. There, it
was pointed out that the holomorphic contribution can emerge
only on account of spontaneous breakdown of the rigid $U(1)$ factor
of the automorphism symmetry of $N=2$ Poincar\'e superalgebra. This
breakdown is induced by a non-zero vacuum value of the physical
scalar field in $N=2$ gauge supermultiplet (corresponding to
a constant superfield strength). Such a constant
value simultaneously generates a central charge
in $N=2$ superalgebra (it is proportional
to the rigid $U(1)$ charge), which, in turn, provides the  hypermultiplet
with a BPS mass. We showed, by straightforward supergraph calculations,
that holomorphic contribution to the off-shell effective action
is the feature inherent just to such a
massive hypermultiplet model. A detailed description of this
theory in harmonic superspace was given
in refs. \cite{10,11,7} (see also \cite{13}).

When calculating the effective action in ref. \cite{5}, we still used the
massless hypermultiplet harmonic superfield propagator and treated
the hypermultiplet mass term as an additional vertex obtained by
shifting the chiral $N=2$ superfield strength by a constant.
Such a procedure of quantum $N=2$ supergraph calculations
led to some technical simplifications.
However, this
way of computation looks rather artificial from
the general point of view. In a massive theory, it is natural
to deal just with the massive propagator and to treat only the
varied part of external superfield as a vertex. Moreover, it
should be pointed out that the case with a non-zero background $N=2$
$U(1)$ superfield strength and the BPS-massive hypermultiplets is
generic for the Coulomb branch of any $N=2$ gauge theory. It is just
$N=2$
supersymmetry with central charges that characterizes the given theory
in the general case, and in order to keep this kind of $N=2$ supersymmetry
manifest at each step of harmonic superspace quantum calculations,
one {\it should} use the appropriate massive propagator for
the hypermultiplet.

The aim of this paper is to work out techniques of calculations
with such {\it massive} hypermultiplet propagator and to demonstrate how
this method leads to a holomorphic contribution to the low-energy
effective action of $N=2$ $U(1)$ gauge superfield. We expect the techniques
given can be, e.g., useful while investigating different aspects of the
effective action in $N=2$ super Yang-Mills theories with spontaneously
broken gauge symmetry.
On the other hand, as was already mentioned, the developing of such
techniques is essential in order to have a well-elaborated quantum harmonic
superspace formalism for the central-charge extended $N=2$
supersymmetry which is inherent in the Coulomb branch of $N=2$ gauge theory.
Note that quantum calculations with the central charge-massive
hypermultiplet propagator were already performed in \cite{7}
to calculate the leading terms in the harmonic superfield effective action
of the hypermultiplet itself.
\vspace{0.5cm}

\noindent
Our starting point will be a model of $q$-hypermultiplet
coupled to an abelian background $N=2$ gauge superfield $V_0^{++}$
of constant strength. The corresponding action reads
\be
\label{1}
S= - \int
{\rm d} \zeta^{(-4)}
\, \breve{q}{}^+ (D^{++} + {\rm i} V_0^{++} )q^+ \;,
\ee
with
\be
\label{3}
V^{++}_0 = - \bar{W}_0 (\theta^{+})^2
- W_0 (\bar \theta^{+})^2 \;,
\ee
$W_0$ being a complex constant.
The hypermultiplet is described by an unconstrained superfield
$q^+ (\zeta)$ and its conjugate $\breve{q}^+ (\zeta)$
defined on the analytic subspace of $N=2$ harmonic superspace
parametrized by
$$
\zeta  \equiv
\{ x_A^m, \theta^{+ \alpha}, \bar \theta^{+ \dot\alpha}, u^{\pm i} \}
$$
with
\bea
& x^m_A = x^m - 2{\rm i} \theta^{(i} \sigma^m {\bar \theta}^{j)}
u^+_i u^-_j \qquad
\theta^{\pm}_{\alpha} = \theta^i_{\alpha} u^{\pm}_i \qquad
\bar \theta^{\pm}_{\dot\alpha} = \bar \theta^i_{\dot\alpha}u^{\pm}_i
\nn \\
& D^{++}q^+  = \Big( u^{+ i}
\displaystyle\frac{\partial}{\partial u^{-i}} -2i\theta^+\sigma^m
\bar \theta^+\displaystyle\frac{\partial}{\partial x^m_A}\Big) q^+ \;.
\nn
\eea
The harmonic variables constrained by
$$
\overline{u^{+ i}} = u^-_i \qquad u^{+i}u^-_i = 1
$$
parametrize the automorphism group $SU(2)_R$ of the $N=2$
Poincar\'e superalgebra.
The integration in \p{1} goes over the analytic superspace.

As was mentioned above, the $\theta $-dependent object
$V^{++}_0 (\zeta)$ can be interpreted as a background $N=2$ analytic
gauge prepotential with a constant strength. Indeed, the standard
expressions for the chiral superfield strength $W$ in terms of the
analytic gauge superfield in the given case yield
\footnote{
We adopt the standard rules of summation over dotted and undotted indices
and use the following notation:
$(D^{\pm})^2=\frac{1}{4}D^{\pm\alpha}D^{\pm}_\alpha$,
$({\bar D}^{\pm})^2=\frac{1}{4}{\bar D}^{\pm}_{\dot\alpha}
{\bar D}^{\pm \dot{\alpha}}$ and $(D^+)^4=(D^+)^2({\bar D}^+)^2$.}
\be
\label{4}
W_0 = -\int{\rm d}u\, (\bar D^{-})^2 V^{++}_0 = {\rm const} \qquad
\bar W_0 = -\int{\rm d}u \,(D^{-})^2 V^{++}_0 = {\rm const}\;.
\ee
The corresponding algebra of covariant derivatives reads
\bea
& \{ {\cal D}^i_\alpha , {\bar {\cal D}}_{\dot{\alpha} j} \} =
-2{\rm i}\, \delta^i_j \, {\cal D}_{\alpha \dot{\alpha}}
\qquad \;\;\;[{\cal D}_{\alpha \dot{\alpha}}, {\cal D}^j_\beta ] =
[{\cal D}_{\alpha \dot{\alpha}}, {\bar {\cal D}}_{\dot{\beta} j} ]=0
\nn \\
& \{ {\cal D}^i_{\alpha},{\cal D}^j_{\beta} \}=
- 2{\rm i}\,\epsilon_{\alpha \beta}\,
\epsilon^{ij} \,{\bar {\bf W}}_0  \qquad
\{ \bar{\cal D}_{\dot\alpha i},\bar{\cal D}_{\dot\beta j} \}=
-2{\rm i}\,\epsilon_{\dot\alpha \dot\beta}\, \epsilon_{ij} \,{\bf W}_0
\label{6}
\eea
where
$$
{\bf W}_0
\left(\begin{array}{l}
q^+ \\
 \breve{q}^+
\end{array} \right)
=
W_0
\left(\begin{array}{l}
\;\;\; q^+ \\
- \breve{q}^+
\end{array} \right) \qquad
{\bar {\bf W}}_0
\left(\begin{array}{l}
q^+ \\
 \breve{q}^+
\end{array} \right)
=
{\bar W}_0
\left(\begin{array}{l}
\;\;\;q^+ \\
- \breve{q}^+
\end{array} \right) \;.
$$
These relations immediately imply that we deal with the central-charge
extended $N=2$ supersymmetry, the central charge being identified with
the generator of global phase $U(1)$ transformations of $q^+ $ which
constitute an obvious symmetry of the action \p{1}. Correspondingly,
the harmonic derivative present in \p{1}
\be
{\cal D}^{++} = D^{++} -{\rm i}\,\left( \bar W_0 (\theta^+)^2 + W_0
(\bar \theta^+)^2 \right)
\ee
has the typical form of the harmonic derivative in the presence of
central charges \cite{8}, under the identification just mentioned.
It is easy to see that \p{1} is invariant with respect to the following
modified $N=2$ supersymmetry transformations
\be
\delta q^+ = - {\rm i} \left(\epsilon Q + \bar \epsilon \bar Q \right)q^+ +
2{\rm i} \left( \epsilon^-\theta^+ \bar W_0 +
\bar\epsilon^-\bar \theta^+ W_0
\right) q^+\;,
\ee
where $\epsilon^{\alpha i}$, $\bar \epsilon^{\dot \alpha}_i $ are $N=2$
supertranslation parameters, $Q, \bar Q$ are the standard $N=2$
supersymmetry
generators in the realization on analytic superfields and
$\epsilon^{\alpha -} = \epsilon^{\alpha i}u^-_i$.
Calculating the Lie bracket of two modified transformations we
find the corresponding generators $\hat Q, \hat{\bar Q}$
to form $N=2$ superalgebra with a central charge proportional to the
rigid $U(1)$ generator
\be
\label{6a}
\{ \hat Q_{\alpha i}, \hat Q_{\beta k} \} = -2{\rm i}\,
\epsilon_{\alpha \beta}\,
\epsilon_{ik}\, \bar{\bf W}_0
\qquad
\{ \hat{\bar Q}_{\dot \alpha i}, \hat{\bar Q}_{\dot\beta k} \}
= 2{\rm i}\,
\epsilon_{\dot\alpha \dot\beta}\,\epsilon_{ik}~{\bf W}_0.
\ee
An equivalent way to introduce such central charges is
to perform the dimensional reduction from $D=6$ harmonic superspace
and to identify a proper combination of the $x^5$- and $x^6$-
translation operators with the rigid $U(1)$ generator (see \cite{10}
for discussion of such a reduction in the harmonic superspace context).
It will become clear soon why we prefer here the interpretation in  terms
of constant background $N=2$ gauge superfield.

It is well known that the presence of the $U(1)$ central charge makes
the hypermultiplet massive \cite{8,5,11}.
This fact directly follows from the dynamical equation
\be
{\cal D}^{++} q^+
\equiv \Big( D^{++} + {\rm i} V^{++}_0 \Big) q^+ = 0
\ee
which implies
\be
(\Box + m^2 ) q^+ = 0 \qquad m = | W_0| \;.
\ee

${}$From the explicit form of $V^{++}_0$ given by eq. (\ref{3})
we see that the automorphism $U(1)_R$--invariance ($R$--symmetry)
\be
\label{7}
\theta^i_\alpha \to {\rm e}^{-{\rm i}\varphi} \theta^i_\alpha \qquad
 \bar\theta^i_{\dot\alpha} \to {\rm e}^{{\rm i}\varphi}
\bar\theta^i_{\dot\alpha}
\ee
of the massless theory ($V^{++}_0 = 0$) is broken in the massive case.
Hence, following the reasoning of ref. \cite{seiberg}, here we have a
potential source for generating a holomorphic effective action at the
quantum level.

The propagator of this theory \cite{10,11,7} satisfies the equation
\be
\label{15}
{\cal D}^{++}G_0^{(1,1)}(\zeta_1, \zeta_2)=
\delta_A^{(3,1)}(\zeta_1, \zeta_2)\; .
\ee
Here
$\delta_A^{(3,1)}(\zeta_1, \zeta_2)$ is
an analytic  $\delta$-function (see \cite{8} for details).
The propagator can be represented as follows
\be
\label{16}
G_0^{(1,1)}=-\displaystyle\frac{1}{\Box +m^2}
(D^+_1)^4(D^+_2)^4 \left\{
\displaystyle\frac{{\rm e}^{{\rm i}(\Omega_0(1)-
\Omega_0(2))} \delta^4(x_1-x_2)\delta^8(\theta_1 -\theta_2)}
{(u^+_1 u^+_2)^3} \right\} \;.
\ee
Here $\Omega = -\bar W_0 \theta^+\theta^- -  W_0\bar\theta^+
\bar\theta^-$
is the `bridge' \cite{8} corresponding to the particular
analytic gauge potential $V^{++}_0$
(see also \cite{11}).
It allows one to choose different representations of the theory.
In  the `$\lambda$-frame' (which we have used throughout
the above consideration)
the harmonic $u^+$-projections of
spinor derivatives contain no central charge
terms,
while the covariant harmonic
derivative is the central charge-extended one ${\cal D}^{++}$.
This frame is distinguished in that the harmonic analyticity can be
made
manifest within it (by choosing $D^+_{\alpha}$ and
$\bar D^+_{\dot\alpha}$
to be `short').
The propagator \p{16}, being analytic, corresponds just
to the  $\lambda $-frame.
Sometimes, it is more convenient to use the `$\tau$-frame'
in which
\be
\label{18}
{\cal D}^+_{\alpha}= e^{-{\rm i}\Omega_0} D^+_{\alpha}
e^{{\rm i}\Omega_0} \qquad
\bar {\cal D}^+_{\dot\alpha}= e^{-{\rm i}\Omega_0}
\bar D^+_{\dot\alpha} e^{{\rm i}\Omega_0} \; .
\ee
Here, ${\cal D}^+_{\alpha}$ and $\bar {\cal D}^+_{\dot\alpha}$ contain
appropriate central charge `connections' and cannot be made `short' by
a change of co-ordinates.
So the harmonic analyticity is covariant in this frame. On the other hand,
the $\tau $-frame covariant harmonic derivative $D^{++}$ contains
no central charge terms.
Surely, the (anti)commutation relations
between different covariant derivatives preserve their form  irrespective
of the choice of the frame  and basis. In particular, the algebra of spinor
derivatives is given by the relations \p{6}.
We will heavily exploit eqs. (\ref{6}) when calculating the
low-energy effective action in the interacting theory.

Now we consider the action of the massive hypermultiplet
in an external gauge abelian superfield $V^{++}_1(\zeta)$
\be
\label{20}
S= - \int {\rm d} \zeta^{(-4)}\,
\breve{q}^+(D^{++}+{\rm i}V^{++}_0 +{\rm i}V^{++}_1)q^+\;.
\ee
This action respects local $U(1)$ symmetry
with analytic gauge parameter and the same central charge-extended
$N=2$ supersymmetry that is inherent in \p{1} ($V^{++}_1$
is assumed to be inert
with respect to the central charge transformations).

Looking at \p{20}, we observe that the action
\p{1} could be arrived at in the following way. One starts
from the standard minimal coupling
of $q^{+}$ to $V^{++}$ with the standard $N=2$ supersymmetry which leaves
invariant separately the free massless $q$-hypermultilet action
and the interaction
term $\sim \breve{q}{}^+V^{++} q^+$.
Then one assumes $V^{++}$
to develop a non-zero vacuum value, $<V^{++}> \equiv V^{++}_0$, and
decompose $V^{++}$ as in \p{20}, $V^{++} = V^{++}_0 + V^{++}_1$,
$<V^{++}_1> = 0$. This decomposition suggests the rearranging of the
free $q$-hypermultiplet action through addition of the $V^{++}_0$--term to
the massless action,
thus giving a mass to $q^+$. The crucial
point is that the decomposition of the original action into the new  free
part and the remainder $\sim \breve{q}{}^+V^{++}_1 q^+$
is invariant under the {\it new} $N=2$ supersymmetry, just the central
charge--extended one described above. This consideration transparently
demonstartes that any $N=2$ gauge theory in the Coulomb branch, i.e. with
a non-zero background value of some gauge superfield $V^{++}$ belonging to
the Cartan subalgebra, respects the $U(1)$ central charge--extended $N=2$
supersymmetry instead of the conventional one.

In accord with the above reasoning, while developing the
perturbation theory in the case at hand, we can proceed
in two different ways. First,
we can regard $V^{++}_0$ as an additional vertex and use
the massless $q^+$ propagator.
Just this way was pursued in ref. \cite{5}.
Secondly, we can regard $V^{++}_0$ as part of the free action
and use the massive propagator. Now we are going to follow this more
natural way. This will allow us, at each step of calculation, to keep
manifest the $N=2$ supersymmetry with $U(1)$ central charge.
\vspace{0.5cm}

\noindent
Similar to ref. \cite{5}, the one-loop effective
action of $V^{++}$ is given as follows
\be
\label{21}
\Gamma[V^{++}]= {\rm i\; Tr\; ln}\;
\big[\delta_A^{(3,1)} +V^{++}_1 G_0^{(1,1)}\big]
\ee
where $G_0^{(1,1)}$ is now the massive propagator (\ref{16}). Let us
consider the perturbation calculation of $\Gamma[V^{++}]$ as a
power series in $V_1^{++}$.

In the second order in $V^{++}_1$, one has
\bea
\label{22}
\Gamma_2[V^{++}_1]&=&-\displaystyle\frac{{\rm i}^3}{2}
\int {\rm d}^4x_1 {\rm d}^4\theta^+_1
{\rm d}u_1 {\rm d}^4x_2 {\rm d}^4\theta^+_2 {\rm d}u_2  \nn\\
&\times&\displaystyle\frac{-1}{\Box_1+m^2}(D_1^+)^4(D_2^+)^4
[{\rm e}^{{\rm i}(\Omega_0(1) -\Omega_0(2))}
\delta^4(x_1-x_2)\delta^8(\theta_1-\theta_2)] \nonumber\\
&\times&\displaystyle\frac{-1}{\Box_2+m^2}(D_2^+)^4(D_1^+)^4
[{\rm e}^{{\rm i}(\Omega_0(2) -\Omega_0(1))}
\delta^4(x_2-x_1)\delta^8(\theta_2-\theta_1)] \nn\\
&\times&\displaystyle\frac{V_1^{++}(x_1,\theta_1,u_1)
V_1^{++}(x_2,\theta_2,u_2)}{(u_1^+u_2^+)^3(u_2^+u_1^+)^3}\;.
\end{eqnarray}
Further calculations are very close to those of \cite{5}, therefore
we write down the final result
\be
\label{23}
\Gamma_2[V^{++}_1]=\frac{{\rm i}^3}{2}\int \frac{{\rm d}^4p}
{(2\pi)^4 (p^2-m^2)^2} \int {\rm d}^4x {\rm d}^4 \theta \; W^2_1
\ee
where $W_1$ is the superfield strength of $V^{++}_1$.

Now we turn to the $n$-th order and consider it in detail.
A straightforward expansion of (\ref{21}) leads to
\bea
\label{24}
\Gamma_n[V^{++}_1]&=&\displaystyle\frac{(-{\rm i})^{n+1}}{n}
\int {\rm d}^4x_1 {\rm d}^4\theta^+_1
{\rm d}u_1 \dots {\rm d}^4x_n {\rm d}^4\theta^+_n {\rm d}u_n \nn\\
&\times&\displaystyle\frac{-1}{\Box_1+m^2}(D_1^+)^4(D_2^+)^4
[{\rm e}^{{\rm i}(\Omega_0(1) - \Omega_0(2))}
\delta^4(x_1-x_2)\delta^8(\theta_1-\theta_2)] \nonumber\\
&\times&\displaystyle\frac{-1}{\Box_2+m^2}(D_2^+)^4(D_3^+)^4
[{\rm e}^{{\rm i}(\Omega_0(2) - \Omega_0(3))}
\delta^4(x_2-x_3)\delta^8(\theta_2-\theta_3)] \nn\\
&\times&
\dots \qquad \dots \qquad \dots \qquad \dots \qquad \dots
\nn\\
&\times&\displaystyle\frac{-1}{\Box_n+m^2}(D_n^+)^4(D_1^+)^4
[{\rm e}^{{\rm i}(\Omega_0(n) - \Omega_0(1))}
\delta^4(x_n-x_1)\delta^8(\theta_n-\theta_1)] \nn\\
&\times&\displaystyle\frac{V_1^{++}(x_1,\theta_1,u_1)
V_1^{++}(x_2,\theta_2,u_2)\dots V_1^{++}(x_n,\theta_n,u_n)}
{(u_1^+u_2^+)^3(u_2^+u_3^+)^3 \dots (u_n^+u_1^+)^3}\;.
\end{eqnarray}
One can introduce the $\tau$-frame derivatives according
to (\ref{18}). Eq. (\ref{24})
then takes the form
\bea
\label{25}
\Gamma_n[V^{++}_1]&=&\displaystyle\frac{(-{\rm i})^{n+1}}{n}
\int {\rm d}^4x_1 {\rm d}^4\theta^+_1
{\rm d}u_1 \dots {\rm d}^4x_n {\rm d}^4\theta^+_n {\rm d}u_n \nn\\
&\times&\displaystyle\frac{-1}{\Box_1+m^2}({\cal D}_1^+)^4({\cal D}_2^+
)^4
[\delta^4(x_1-x_2)\delta^8(\theta_1-\theta_2)] \nonumber\\
&\times&\displaystyle\frac{-1}{\Box_2+m^2}({\cal D}_2^+)^4({\cal D}_3^+
)^4
[\delta^4(x_2-x_3)\delta^8(\theta_2-\theta_3)] \nn\\
&\times&\dots \qquad \dots \qquad \dots \qquad \dots \qquad \dots
 \nn\\
&\times&\displaystyle\frac{-1}{\Box_n+m^2}({\cal D}_n^+)^4({\cal D}_1^+
)^4
[\delta^4(x_n-x_1)\delta^8(\theta_n-\theta_1)] \nn\\
&\times&\displaystyle\frac{V_1^{++}(x_1,\theta_1,u_1)
V_1^{++}(x_2,\theta_2,u_2)\dots V_1^{++}(x_n,\theta_n,u_n)}
{(u_1^+u_2^+)^3(u_2^+u_3^+)^3 \dots (u_n^+u_1^+)^3}\;.
\end{eqnarray}
After restoring the full Grassmann integration measure, integrating  over
$\theta_3, \dots ,\theta_n$ and passing to the momentum space
we obtain
\bea
\label{26}
&&\Gamma_n[V^{++}_1]=\displaystyle\frac{(-{\rm i})^{n+1}}{n}
\int\displaystyle\frac{{\rm d}^4p_1 \dots {\rm d}^4p_n
{\rm d}^8 \theta_1 {\rm d}^8 \theta_2
{\rm d}u_1 \dots {\rm d}u_1 \dots {\rm d} u_2}
{ (2\pi)^{4n} (p_1^2 -m^2) \dots (p_n^2 -m^2)} \nn\\
&&\times\displaystyle\frac{\delta^8(\theta_1-\theta_2)V^{++}_1
(\theta_1,u_1)
V^{++}_1(\theta_2,u_2)}{(u^+_1 u^+_2)^3(u^+_2 u^+_3)^3
\dots (u^+_n u^+_1)^3}
[{\cal D}_1^+(\theta_1)]^4 \{ V^{++}_1(\theta_1,u_n)
[{\cal D}_2^+(\theta_2)]^4  \nn\\
&&\times\{ V^{++}_1(\theta_2,u_3)
[{\cal D}_3^+(\theta_2)]^4 \{ V^{++}_1(\theta_2,u_4)\dots
[{\cal D}_{n-1}^+(\theta_2)]^4  [{\cal D}_n^+(\theta_1)]^4
\delta^8(\theta_2-\theta_1) \}\dots\}\}\;.
\eea
Here the dependence on momenta is suppressed.
The only terms which contribute in the local limit are
\bea
\label{27}
\Gamma_n[V^{++}_1]&=&\displaystyle\frac{(-{\rm i})^{n+1}}{n}
\int\displaystyle\frac{{\rm d}^4p_1 \dots {\rm d}^4p_n
{\rm d}^8 \theta_1 {\rm d}^8 \theta_2
{\rm d}u_1 \dots {\rm d}u_1 \dots {\rm d}u_2}
{ (2\pi)^{4n} (p_1^2 -m^2) \dots (p_n^2 -m^2)} \nn\\
&\times&\displaystyle\frac{\delta^8(\theta_1-\theta_2)V^{++}_1
(\theta_1,u_1)
V^{++}_1(\theta_2,u_2)}{(u^+_1 u^+_2)^3(u^+_2 u^+_3)^3
\dots (u^+_n u^+_1)^3}
[\bar{\cal D}_1^+(\theta_1)]^2  V^{++}_1(\theta_1,u_n) \nn\\
&\times&[\bar{\cal D}_2^+(\theta_2)]^2  V^{++}_1(\theta_2,u_3)
[\bar{\cal D}_3^+(\theta_2)]^2  V^{++}_1(\theta_2,u_4)
\dots [\bar{\cal D}_{n-2}^+(\theta_2)]^2  V^{++}_1(\theta_2,u_{n-1})
\nn\\
&\times&[{\cal D}_2^+(\theta_2)]^2 [{\cal D}_3^+(\theta_2)]^2
\dots [{\cal D}_n^+(\theta_2)]^2 [{\cal D}_1^+(\theta_1)]^2
[\bar{\cal D}_{n-1}^+(\theta_2)]^2 [\bar{\cal D}_{n}^+(\theta_2)]^2
\delta^8(\theta_2-\theta_1)\nn\\
&+& \qquad{\rm c.c.}
\eea
These terms lead to the holomorphic and antiholomorphic contributions
in the
low-energy limit. Note the relation
${\cal D}^+V_1^{++}=D^+V^{++}_1$ which follows from the fact
that $V^{++}_1$ is invariant under global $U(1)$ and, hence, under the
central charge.

We will expand the spinor derivatives acting on a given superfield
over those harmonics on which this superfield depends.
For example, let us consider $(\bar D^+)^2V^{++}_1(u_n)$
\be
\label{28}
\bar D^+_{1 \dot\alpha} =\bar D^{i}_{\dot\alpha}u^+_{1 i} =
\bar D^-_{n \dot\alpha}(u_n^+ u^+_1) -
\bar D^+_{n \dot\alpha}(u_n^- u^+_1)\;.
\ee
Due to the analyticity of $V^{++}_1$, only one term is essential here
\be
\label{29}
(\bar D^+_1)^2 V^{++}_1(u_n)= (\bar D^-_n)^2 V^{++}_1(u_n)(u^+_1 u^+_n)^2~.
\ee
Then, equation (\ref{27}) can be rewritten as
\bea
\label{30}
\Gamma_n[V^{++}_1]&=&\displaystyle\frac{(-{\rm i})^{n+1}}{n}
\int\displaystyle\frac{{\rm d}^4p_1 \dots {\rm d}^4p_n
{\rm d}^8 \theta_1 {\rm d}^8 \theta_2
{\rm d}u_1 \dots {\rm d}u_1 \dots {\rm d}u_2}
{ (2\pi)^{4n} (p_1^2 -m^2) \dots (p_n^2 -m^2)} \nn\\
&\times&\displaystyle\frac{\delta^8(\theta_1-\theta_2)V^{++}_1
(\theta_1,u_1)
V^{++}_1(\theta_2,u_2)}
{(u^+_1 u^+_2)^3(u^+_2 u^+_3)\dots (u^+_{n-2} u^+_{n-1})
(u^+_{n-1} u^+_{n})^3(u^+_{n} u^+_{1})^3} \nn\\
&\times&[\bar D_3^-(\theta_2)]^2  V^{++}_1(\theta_2,u_3)
[\bar D_4^-(\theta_2)]^2  V^{++}_1(\theta_2,u_4)
\dots [\bar D_n^+(\theta_2)]^2  V^{++}_1(\theta_2,u_n) \nn\\
&\times&[{\cal D}_2^+(\theta_2)]^2 [{\cal D}_3^+(\theta_2)]^2
\dots [{\cal D}_n^+(\theta_2)]^2 [{\cal D}_1^+(\theta_1)]^2
[\bar{\cal D}_{n-1}^+(\theta_2)]^2 [\bar{\cal D}_{n}^+(\theta_2)]^2
\delta^8(\theta_2-\theta_1) \nn\\
&+& \qquad{\rm c.c.}
\eea
In the chain of spinor derivatives acting on the $\delta$-function
the derivatives $\bar {\cal D}^+$ can be eliminated using the
identity
$$
\bar \delta(\bar \theta_1 -\bar\theta_2)
(\bar {\cal D}^+_{n-1})^2 (\bar {\cal D}^+_{n})^2
\bar \delta(\bar \theta_2 -\bar\theta_1)=(u^+_{n-1}u^+_n)^2
\bar \delta(\bar \theta_2 -\bar\theta_1)~.
$$
Respectively, we should not care about $[{\cal D}^+_1(\theta_1)]^2$
since it depends on $\theta_1$, whereas the other blocks depend
on $\theta_2$.

Let us write the chain of derivatives in the following form
\bea
\label{31.5}
[{\cal D}_2^+(\theta_2)]^2 [{\cal D}_3^+(\theta_2)]^2
& \dots & [{\cal D}_n^+(\theta_2)]^2=
\displaystyle\frac{1}{4^{n-1}}
({\cal D}_2^+ {\cal D}_{2}^+)
({\cal D}_3^{+} {\cal D}_{3}^+)\dots
({\cal D}_n^{+} {\cal D}_{n}^+)\nn\\
& = &\displaystyle\frac{(-1)^{n-2}}{4^{n-1}}
(u^+_2 u^+_3)(u^+_3 u^+_4)\dots (u^+_{n-2} u^+_{n-1})(u^+_{n-1} u^+_n)
\nn \\
& \times & ({\cal D}_2^{+} {\cal D}_{3}^-)
({\cal D}_3^{+} {\cal D}_{4}^-)\dots
({\cal D}_{n-2}^{+} {\cal D}_{n-1}^-)
({\cal D}_{n-1}^{+} {\cal D}_{n}^-)
({\cal D}_n^{+} {\cal D}_{n}^+)\;.
\eea
As the next step, we expand the derivative
${\cal D}_{n-1}^{+ \alpha}$ in \p{31.5}
over the $n$-th set of harmonics. The result reads
\bea
\label{32}
[{\cal D}_2^+(\theta_2)]^2 [{\cal D}_3^+(\theta_2)]^2
 \dots  [{\cal D}_n^+(\theta_2)]^2=
\displaystyle\frac{(-1)^{n-2}}{4^{n-1}}
(u^+_2 u^+_3)(u^+_3 u^+_4)\dots (u^+_{n-1} u^+_n) \nn\\
 \times  \Big\{ ({\cal D}_2^{+} {\cal D}_{3}^-)
({\cal D}_3^{+} {\cal D}_{4}^-)\dots
({\cal D}_{n-2}^{+} {\cal D}_{n-1}^-)
({\cal D}_{n}^{-} {\cal D}_{n}^-)
({\cal D}_n^{+} {\cal D}_{n}^+)(u^+_n u^+_
{n-1})  \nn\\
 -  ({\cal D}_2^{+} {\cal D}_{3}^-)
({\cal D}_3^{+} {\cal D}_{4}^-)\dots
({\cal D}_{n-2}^{+} {\cal D}_{n-1}^-)
({\cal D}_{n}^{+} {\cal D}_{n}^-)
({\cal D}_n^{+} {\cal D}_{n}^+)(u^-_n u^+_{n-1}) \Big\}\;.
\eea
Further, in the first term we expand ${\cal D}^-_{n-1}$ again in terms
of the $n$th harmonics (we suppress spinor indices).
Only the ${\cal D}^+_{n}$ projection
survives,
then we commute it with ${\cal D}^-_{n}$ using the
algebra (\ref{6}). Since $({\cal D}^+)^3=0$, only the commutator remains.
The situation is simpler with the second term. We just anticommute
${\cal D}^{+}_{n}$ with ${\cal D}^-_{n}$  and
then expand ${\cal D}^-_{n-1}$ over the $n$-th harmonics.
We arrive at
\bea
\label{33}
[{\cal D}_2^+(\theta_2)]^2 [{\cal D}_3^+(\theta_2)]^2
\dots [{\cal D}_n^+(\theta_2)]^2 &=&
\displaystyle\frac{(-1)^{n-2}}{4^{n-1}} 4{\rm i}\bar{W_0}
(u^+_2 u^+_3)(u^+_3 u^+_4)\dots (u^+_{n-1} u^+_n) \nn\\
& \times &({\cal D}^{+}_2 {\cal D}^-_{3}) \dots
({\cal D}^{+}_{n-3} {\cal D}^-_{n-2})
({\cal D}^{+}_{n-2} {\cal D}^-_{n})
({\cal D}^{+}_{n} {\cal D}^-_{n}) \nn \\
& \times & \big\{ (u^+_n u^+_{n-1})(u^-_n u^-_{n-1})- (u^+_n u^-_{n-1})
(u^-_n u^+_{n-1}) \big\}\;.
\eea
Here the harmonic expression in the
last line is equal to 1 \cite{8}
and we finally get
\bea
\label{34}
[{\cal D}_2^+(\theta_2)]^2 [{\cal D}_3^+(\theta_2)]^2
\dots [{\cal D}_n^+(\theta_2)]^2 & = &
\displaystyle\frac{(-1)^{n-2}}{4^{n-1}} 4{\rm i}\bar W_0
(u^+_2 u^+_3)(u^+_3 u^+_4)\dots (u^+_{n-1} u^+_n) \nn\\
& \times &({\cal D}^{+}_2 {\cal D}^-_{3}) \dots
({\cal D}^{+}_{n-3} {\cal D}^-_{n-2})
({\cal D}^{+}_{n-2} {\cal D}^-_{n})
({\cal D}^{+}_{n} {\cal D}^-_{n})\;.
\eea

Comparing \p{34} with \p{31.5}, we observe that the former can be formally
obtained from the latter by replacing the block
${\cal D}^-_{ n-1 \alpha}{\cal D}^{+ \beta}_{n-1}$
by $4{\rm i}\delta^{\beta}_{\alpha}$.  This observation allows the
process to go on. The chain $[{\cal D}_2^+(\theta_2)]^2 [{\cal
D}_3^+(\theta_2)]^2 \dots [{\cal D}_n^+(\theta_2)]^2$ is finally
reduced to
$$
\displaystyle\frac{(-{\rm i})^{n-2}}{4} \bar{W_0}^{n-2}
(u^+_2 u^+_3)\dots (u^+_{n-1} u^+_n) ({\cal D}^{+}_2
{\cal D}^+_{n})\;.
$$
Since
$$
\delta(\theta_1 -\theta_2)
\displaystyle\frac{1}{4}
({\cal D}^+_{2}) ({\cal D}^+_{n})   ({\cal D}^+_{1})^2
\delta(\theta_2 -\theta_1)=-(u^+_{1}u^+_2)(u^+_{n}u^+_1)
\delta(\theta_2 -\theta_1)
$$
we can integrate over $\theta_2$ and obtain the following expression
\bea
\label{36}
\Gamma_n[V^{++}_1]=\displaystyle\frac{(-{\rm i})^{n-2}}{n}
\int\displaystyle\frac
{{\rm d}^4p_1 \dots {\rm d}^4p_n {\rm d}^8 \theta {\rm d}u_1 \dots
{\rm d}u_n}{(2\pi)^4 (p_1-m^2)^2 \dots (p_n-m^2)^2} \nn \\
\times \bar{W}_0^{n-2}\displaystyle\frac{V_1^{++}(1)V_1^{++}(2)}
{(u^+_1 u^+_2)^2}
(\bar{D}_3^-)^2V^{++}_1(3) \dots (\bar{D}_n^-)^2V^{++}_1(n) +{\rm
c.c.}
\eea

Putting together all the above, we get in the local limit the
holomorphic and antiholomorphic contributions
\be
\label{37}
\Gamma_n[V^{++}_1]=-\frac{{\rm i}}{n} \int\frac{{\rm d}^4p}{(2\pi)^4
(p^2-W_0 {\bar W}_0)^n}\int {\rm d}^4x {\rm d}^4\theta \;
\bar W_0^{n-2} W_1^n
+{\rm c.c.}
\ee
where we have accounted the relation $m^2 = W_0 \bar W_0$.
The momentum integral here is automatically infrared-finite;
for massive particles there is no essential difference between
the Wilsonian and effective actions.

The total low-energy effective action is the sum of $\Gamma_n$:
\be
\label{39}
\Gamma[V^{++}]=\sum_{n=2}^{\infty} \Gamma_n[V^{++}_1]\;.
\ee
After doing the sum we end up with the following final result
for the renormalized low-energy effective action
\be
\label{40}
\Gamma[V^{++}]=-\frac{1}{64\pi^2}\int {\rm d}^4x {\rm d}^4\theta\;
W^2\; {\rm ln}\, \frac{W^2}{M^2} + {\rm c.c.}\, ,
\qquad W=W_0 + W_1\,,
\ee
with $M$ being a normalization scale. Expression (\ref{40})
coincides with that obtained in ref. \cite{5} by making use of the
massless $q^+$ propagator.

As was argued in \cite{5}, the holomorphic contribution to
low-energy effective action can emerge only on account of spontaneous
breakdown of the rigid $U(1)_R$ symmetry (in accord with the reasoning
of ref. \cite{seiberg}).
If this symmetry is unbroken, there is no room for holomorphic
corrections. Therefore, such corrections should vanish in the limit
$W_0 \rightarrow 0$ corresponding to the massless hypermultiplet.
To get the low-energy effective action in massless limit, we have, from the
very beginning, to work with the Wilsonian action obtained
by cutting off the momentum integrals entering the quantum corrections
(\ref{37}) at some low-energy scale $\Lambda$.
The massless limit corresponds to $W_0 \rightarrow 0$ while keeping
$\Lambda$ fixed.  Then, from (\ref{37}) one observes that for $n>2$
all corrections $\Gamma_n$ vanish in the massless limit; on the other
hand, $\Gamma_2$ can be removed by a renormalization. This is
in complete agreement with the results of ref. \cite{5} and confirms  our
statement that the holomorphic effective action is entirely due
to non-zero central charge.
\vspace{0.5cm}

\noindent
Let us summarize the results. We have described the massive
hypermultiplet with the mass generated via coupling to the abelian
gauge superfield with a constant strength. This coupling breaks
$U(1)_R$-automorphism
symmetry of $N=2$ superalgebra and simultaneously leads to the
central charge. When calculating the low-energy one-loop effective action
of an external $N=2$ gauge multiplet,
we used the manifestly $N=2$ supersymmetric techniques of quantum
computations with the massive hypermultiplet propagator and
found the leading terms in the one-loop low-energy effective action
to be holomorphic. This result coincides with that given in ref. \cite{5}
where it was obtained by using the massless
propagator and considering the $U(1)_R$ breaking term as part of the
vertex. The holomorphic contribution has a structure
analogous to the one found in ref. \cite{seiberg} where it was pointed out
that eq. (\ref{40}) is the only expression possible which
reproduces the $U(1)_R$-anomaly.

Our manifestly $N=2$ supersymmetric
off-shell calculation clearly demonstrates that such a
holomorphic contribution is an entire effect of non-zero
$U(1)$ central charge in $N=2$ superalgebra. This central charge
measures the breaking of the $U(1)_R$-automorphism symmetry, thus
establishing a link with the reasoning of ref. \cite{seiberg}.
Clearly, the holomorphic contributions of the same structure should
emerge in any model where some abelian $V^{++}$ couples to
superfields with non-trivial $U(1)$ central charge, the specificity of
the given coupling being encoded in the numerical coefficient with
which the holomorphic contribution enters. In particular, this
phenomenon occurs in the pure $SU(2)$ $N=2$ super Yang-Mills
theory with a non-zero $<V^{++}_3>$ giving rise to a non-trivial  $U(1)$
central charge and the related BPS masses for the non-diagonal
components of the $SU(2)$ algebra valued $V^{++}$. Recall that the
consideration of ref. \cite{seiberg} referred just to such a
non-abelian situation.

\vspace{1cm}

\noindent
{\bf Acknowledgements.}
The authors are grateful to B.A. Ovrut and B.M. Zupnik for valuable
discussions. This work was partially supported by INTAS grant,
INTAS-96-0308. I.L.B., E.A.I. and S.M.K. acknowledge a partial support
from RFBR-DFG grant, project No 96-0200180. I.L.B. and S.M.K. are indebted
to RFBR grant, project No 96-02-1607 for partial support. E.A.I.
acknowledges a partial support from RFBR grant, project No 96-02-17634,
and INTAS grants, INTAS-93-127-ext and INTAS-96-0538.

\end{document}